\documentclass[doublecol,a4]{epl2}
\newcommand{\be}{\begin{equation}}\newcommand{\ee}{\end{equation}}
\newcommand{\bea}{\begin{eqnarray}}\newcommand{\eea}{\end{eqnarray}}
\newcommand{\brr}{\begin{array}}\newcommand{\err}{\end{array}}
\newcommand{\bit}{\begin{itemize}}\newcommand{\eit}{\end{itemize}}
\newcommand{\ben}{\begin{enumerate}}\newcommand{\een}{\end{enumerate}}

\newcommand{\ba}{\begin{array}}
\newcommand{\ea}{\end{array}}

\def\lan{\langle}
\def\lf{\left}

\def\ran{\rangle}

\def\ri{\right}

\def\de{\delta}

\def\1{{_{1}}}\def\2{{_{2}}}

\newcommand{\ide}{1\hspace{-1mm}{\rm I}}

\def\noHe0{:\;\!\!\;\!\!:H_e(0):\;\!\!\;\!\!:}
\def\noHm0{:\;\!\!\;\!\!:H_\mu(0):\;\!\!\;\!\!:}

\title{Entanglement in neutrino oscillations}
\shorttitle{Entanglement in neutrino oscillations}
\author{Massimo Blasone\inst{1,2} \and Fabio Dell'Anno\inst{1,2,3} \and Silvio De Siena\inst{1,2,3}
\and Fabrizio Illuminati \inst{1,2,3,4}}
\shortauthor{M.~Blasone, F.~Dell'Anno, S.~De~Siena and F.~Illuminati}

\institute{
  \inst{1} Dipartimento di Matematica e Informatica,
Universit\`a degli Studi di Salerno, Via Ponte don Melillo,
I-84084 Fisciano (SA), Italy
\\
  \inst{2} INFN Sezione di Napoli,
Gruppo collegato di Salerno, Baronissi (SA), Italy
\\
\inst{3} CNR-INFM Coherentia, Napoli, Italy
\\
\inst{4}
ISI
Foundation for Scientific Interchange, Viale Settimio Severo 65,
I-10133 Torino, Italy
}


\abstract{Flavor oscillations in elementary particle physics
are related to multi-mode entanglement of single-particle states.
We show that mode entanglement can be expressed in terms of flavor transition
probabilities, and therefore that single-particle entangled states acquire
a precise operational characterization in the context of particle mixing.
We treat in detail the physically relevant cases of two- and three-flavor
neutrino oscillations, including the effective measure of $CP$ violation.
We discuss experimental schemes for the transfer of the quantum information
encoded in single-neutrino states to spatially delocalized two-flavor charged lepton
states, thus showing, at least in principle, that single-particle entangled states
of neutrino mixing are legitimate physical resources for quantum information tasks.}

\begin{document}

\maketitle

Various branches of condensed matter, atomic physics, and quantum
optics have evolved in recent years towards the investigation and
development of schemes for quantum information and computation
science \cite{NielsenChuang}. To this aim, entanglement is a key
ingredient and a crucial physical resource. Different forms of
entanglement have been proven to be equivalent to observable
quantifiers of performance success in quantum information protocols
either with discrete \cite{Horodecki} or continuous variables
\cite{AdessoIlluminatiPRL2005}. In the present work, we investigate
the operational meaning of entanglement in the context of elementary
particles physics. We will show that single-particle (mode)
entanglement associated to particle mixing can be expressed in terms of transition probabilities
in flavor oscillations, and can be exploited for quantum informational tasks.

The concept of single-particle entanglement has been introduced and
elucidated in a recent series of important theoretical papers
\cite{Zanardi,VanEnk,Vedral}. Its use has been discussed in various
contexts of quantum information, including teleportation, quantum
cryptography, and violation of Bell inequalities \cite{Bjork,Lee},
and later experimentally demonstrated with single-photon systems
\cite{Lombardi,Lvovski,Hessmo}. Existing schemes to probe
non-locality in single-photon states have been subsequently
generalized to include massive particles \cite{Vedral2}. In the
present work we extend the discussion to the arena of elementary
particles and provide a general operational characterization of
single-particle entanglement in this context by showing its
connection with the transition probabilities in any quantum system
oscillating between different modes. We then show how this form of
entanglement is in principle a real physical resource for the
realization of quantum information protocols by discussing explicit
experimental schemes for transferring it to spatially separated
modes of stable leptonic particles. These results allow to place
mode entanglement in neutrino oscillations on equal footing with
mode entanglement in single-particle atomic and optical systems.
Furthermore, we show how the single-particle entanglement quantifies
$CP$ violation in neutrino mixing.

Flavor mixing of neutrinos for three generations is described by the
$3 \times 3$ {Pontecorvo-Maki-Nakagawa-Sakata (PMNS)}
unitary mixing matrix $\mathbf{U}(\tilde{\theta},\delta)$
\cite{Cheng-Li},
\begin{widetext}
\begin{equation}
\mathbf{U}(\tilde{\theta},\delta)\,=\,\lf(\ba{ccc}
c_{12}c_{13} & s_{12}c_{13} & s_{13}e^{-i\de} \\
-s_{12}c_{23}-c_{12}s_{23}s_{13}e^{i\de} &
c_{12}c_{23}-s_{12}s_{23}s_{13}e^{i\de} & s_{23}c_{13} \\
s_{12}s_{23}-c_{12}c_{23}s_{13}e^{i\de} &
-c_{12}s_{23}-s_{12}c_{23}s_{13}e^{i\de} & c_{23}c_{13} \ea\ri)\,
\,,
\label{CKMmatrix}
\end{equation}
\end{widetext}
where
$(\tilde{\theta},\delta)\equiv(\theta_{12},\theta_{13},\theta_{23};\delta)$
and
$c_{ij}\equiv\cos\theta_{ij}$,  $s_{ij}\equiv\sin\theta_{ij}$.
The parameters $\theta_{ij}$ are the mixing angles, and $\delta$ is
the phase responsible for $CP$ violation.
{Here, without loss of generality we consider only Dirac neutrinos.
In the instance of Majorana neutrinos,
two additional CP-violating phases are present,
which, however, as it is well known, do not affect the
physics of neutrino oscillations.}
The three-flavor neutrino states are defined as
\begin{equation}
|\underline{\nu}^{(f)}\ran \,=\, \mathbf{U}(\tilde{\theta},\delta)
\, |\underline{\nu}^{(m)}\ran
\label{fermix3}
\end{equation}
where $|\underline{\nu}^{(f)}\ran \,=\, \left(
|\nu_e\ran,|\nu_\mu\ran, |\nu_\tau\ran \right)^{T}$ are the states
with definite flavor and $|\underline{\nu}^{(m)}\ran \,=\, \left(
|\nu_1\ran,|\nu_2\ran, |\nu_3\ran \right)^{T}$ those with definite
mass. Let us recall that both $|\nu_{\alpha}\rangle$
$(\alpha=e,\mu,\tau)$ and $|\nu_{j}\rangle$  $(j=1,2,3)$ are
orthonormal, i.e. $\langle \nu_{\alpha}|\nu_{\beta}\rangle =
\delta_{\alpha,\beta}$ and $\langle \nu_{j}|\nu_{k}\rangle =
\delta_{j,k}$.

Neutrino oscillations are due to neutrino mixing and
neutrino mass differences. The neutrino states $|\nu_j\rangle$ have
definite masses $m_{j}$ and definite energies $E_{j}$.  Their
propagation can be described by plane wave solutions of the form
$|\nu_j(t)\rangle = e^{-i E_{j}t} |\nu_{j}\rangle$. The time
evolution of the flavor neutrino states is given by:
\begin{eqnarray} \nonumber
|\underline{\nu}^{(f)}(t)\rangle & =& \mathbf{\widetilde{U}}(t)
|\underline{\nu}^{(f)}\rangle \,,
\\ [2mm]
\mathbf{\widetilde{U}}(t) & \equiv &
\mathbf{U}(\tilde{\theta},\delta) \, \mathbf{U}_{0}(t) \,
\mathbf{U}(\tilde{\theta},\delta)^{-1} \,,
\label{flavstateevolution}
\end{eqnarray}
where $|\underline{\nu}^{(f)}\rangle$ are the flavor states at
$t=0$, $\mathbf{U}_{0}(t) = diag (e^{-i E_{1}t},e^{-i E_{2}t},e^{-i
E_{3}t})$, and $\mathbf{\widetilde{U}}(t=0)=\ide $. At time $t$ the
transition probability for $\nu_{\alpha}\rightarrow\nu_{\beta}$ is
\begin{equation}
P_{\nu_{\alpha}\rightarrow\nu_{\beta}}(t) \,=\,
|\langle\nu_{\beta}|\nu_{\alpha}(t)\rangle|^{2}
\,=\,|\mathbf{\widetilde{U}}_{\alpha \beta}(t)|^{2} \, ,
\label{neutrinooscillation}
\end{equation}
where $\alpha,\beta = e, \mu, \tau \,.$ The transition probability
$P_{\nu_{\alpha}\rightarrow\nu_{\beta}}(t)$ is a function of the
energy differences $\Delta E_{jk} = E_{j}-E_{k}$ $(j,k=1,2,3)$ and
of the mixing angles. Since {the current neutrino
experiments deal with ultra-relativistic neutrinos}, the standard
adopted approximation is $\Delta E_{jk} \simeq \frac{\Delta
m_{jk}^{2}}{2E}$, where $\Delta m_{jk}^{2} = m_{j}^{2}-m_{k}^{2}$
and $E=|\overrightarrow{p}|$ is the energy of a massless neutrino
(all massive neutrinos are assumed to have the same momentum
$\overrightarrow{p}$).

Flavor neutrinos are identified via charged current weak interaction
processes, together with their associated charged leptons.
{ In the Standard Model (SM),
where neutrinos are taken to be massless,
%the form of the interaction is such that the
flavor is strictly conserved in such processes.  On the other hand,
neutrino mixing, consisting in a mismatch between flavor and mass,
is at the basis of neutrino oscillations and CP violation.
The introduction of neutrino masses
as a correction to the SM is a necessary
condition to explain such effects.}

When neutrino mixing is considered,
%flavor neutrino fields are superpositions of neutrino
%fields with definite masses and
{loop corrections produce violations of lepton flavor in the charged
current vertices: however,
these effects are extremely small and essentially vanish in the
relativistic limit\cite{Casas}}. Consequently, neutrino states entering weak
interaction processes, like the ones where flavor neutrinos are
created or detected, must be eigenstates of flavor neutrino charges.
The corresponding operators can be rigorously defined together with
their eigenstates in the context of Quantum Field Theory (QFT)
\cite{Blasone}. In the relativistic limit, the exact QFT flavor
states reduce to the usual Pontecorvo flavor states, which define
the flavor modes as legitimate and physically well-defined
individual entities. Mode entanglement can thus be defined and
studied in analogy with the static case
\cite{StaticEntanglement}.

Equipped with the above setting, one can establish the
following correspondence with three-qubit states:
$|\nu_{e}\rangle
\equiv |1\rangle_{\nu_{e}} |0\rangle_{\nu_{\mu}}
|0\rangle_{\nu_{\tau}}$, $|\nu_{\mu}\rangle \equiv
|0\rangle_{\nu_{e}} |1\rangle_{\nu_{\mu}} |0\rangle_{\nu_{\tau}}$,
$|\nu_{\tau}\rangle \equiv |0\rangle_{\nu_{e}} |0\rangle_{\nu_{\mu}}
|1\rangle_{\nu_{\tau}}$. States $|0\rangle_{\nu_{\alpha}}$ and
$|1\rangle_{\nu_{\alpha}}$ correspond, respectively, to the absence
and the presence of a neutrino in mode $\alpha$. Entanglement is
thus established among flavor modes, in a single-particle setting.
Eq.~(\ref{flavstateevolution}) can then be recast as
\begin{eqnarray}
|\nu_{\alpha}(t)\rangle &=& \mathbf{\widetilde{U}}_{\alpha e}(t)
|1\rangle_{\nu_{e}} |0\rangle_{\nu_{\mu}} |0\rangle_{\nu_{\tau}} +
\mathbf{\widetilde{U}}_{\alpha \mu}(t) |0\rangle_{\nu_{e}} |1\rangle_{\nu_{\mu}}
|0\rangle_{\nu_{\tau}} \nonumber \\ & & + \, \mathbf{\widetilde{U}}_{\alpha
\tau}(t) |0\rangle_{\nu_{e}} |0\rangle_{\nu_{\mu}} |1\rangle_{\nu_{\tau}} \, ,
\label{flavorWstate}
\end{eqnarray}
with the normalization condition
$\sum_{\beta}|\mathbf{\widetilde{U}}_{\alpha \beta}(t)|^{2}=1$
$(\alpha,\beta=e,\mu,\tau)$. The time-evolved states
$|\underline{\nu}^{(f)}(t)\rangle$ are entangled superpositions of
the three flavor eigenstates with time-dependent coefficients.
{It is important to remark that, although dealing with similar physical systems,
both the framework and the aim of the present paper differ substantially from those
of Ref.~\cite{StaticEntanglement}.
In the latter, by exploiting the wave packet approach,
the multipartite entanglement, associated with the
multiqubit space of mass modes, has been analyzed in connection with
the ``decoherence'' effects induced by free evolution.
In the present work, by exploiting the plane-wave approximation,
the entanglement is quantified with respect to the multiqubit space
associated with flavor modes, and is related to the quantum information
encoded in the neutrino flavor states, which is in principle experimentally
accessible, as we will show by devising an explicit scheme for the information
transfer.}

States of the form Eq.~(\ref{flavorWstate}) belong to the class of
$W$ states. These, together with the $GHZ$ states, define the two
possible sets of states with tripartite entanglement that are
inequivalent under local operations and classical communication
\cite{DurrVidalCirac}. In some instances, only two neutrinos are
significantly involved in the mixing. For example, only the
transition $\nu_{\mu}\leftrightarrow\nu_{\tau}$ is relevant for
atmospheric neutrinos, while only the transitions of the type
$\nu_{e}\leftrightarrow\nu_{\alpha}$ are relevant for solar
neutrinos. For two-flavor mixing the mixing matrix
 $\mathbf{U}(\tilde{\theta},\delta)$ reduces to the $2 \times 2$
rotation matrix $\mathbf{U}(\theta)$,
\begin{equation}
\mathbf{U}(\theta) = \left( \begin{array}{cc}
  \cos\theta & \sin\theta \\
  -\sin\theta & \cos\theta
\end{array}
\right) \,,
\end{equation}
the evolution operator reads $\mathbf{U}_{0}(t) = diag (e^{-i
E_{1}t},e^{-i E_{2}t})$, and the time-evolved flavor states yield
the Bell-like superposition ($\alpha=e,\mu$):
\begin{equation}
|\nu_{\alpha}(t)\rangle \,=\, \mathbf{\widetilde{U}}_{\alpha e}(t)
|1\rangle_{\nu_{e}} |0\rangle_{\nu_{\mu}} + \mathbf{\widetilde{U}}_{\alpha \mu}(t)
|0\rangle_{\nu_{e}} |1\rangle_{\nu_{\mu}}.
\label{flavorBellstate}
\end{equation}
%%%%%%%%%%%%%%%%%%%%%%%%%%%%%%%%%%%%%%%%%%%%%%%%%%%%%%%%%%%%%%%%%%%%%%%%%%%%%%%%%%%%%%%%%%%%%%%%%%%%%%%%%%%%%

Bipartite entanglement of pure states is unambiguously quantified by
the von Neumann entropy or by any other monotonic function of the
former \cite{EntRevHorodecki}. Among entanglement monotones, the
linear entropy has a special physical significance because it is
directly linked to the purity of the reduced states, and enters in
the fundamental monogamy inequalities for distributed entanglement
in the multipartite setting \cite{EntRevHorodecki}. As one moves
from the two- to the three-flavor scenario, multipartite
entanglement measures are readily available in terms of functions of
bipartite measures \cite{Wallach,Brennen,Oliveira}. Representative
of this type of measures is the global entanglement. It is defined
as the sum of all the two-qubit entanglements between a single
subsystem and each of the remaining ones \cite{Wallach}, and can be
expressed as the average subsystem linear entropy \cite{Brennen}.
Global entanglement can then be generalized by constructing the set
of mean linear entropies associated to all possible bi-partitions of
the entire system \cite{Oliveira}.
An alternative characterization of multipartite entanglement is given in Refs.\cite{WeiGold,geomet}.

 Let $\rho=|\psi\ran\lan \psi|$ be the density operator
corresponding to a pure state $|\psi\ran$, describing the system $S$
partitioned into $N$ parties. Consider the bipartition of the
$N$-partite system $S=\{S_{1},S_{2},\ldots,S_{N}\}$ in two
subsystems $S_{A_{n}}=\{S_{i_{1}},S_{i_{2}},\ldots,S_{i_{n}}\}$,
with $1\leq i_{1}<i_{2}<\ldots <i_{n}\leq N$ $(1\leq n <N)$, and
$S_{B_{N-n}}=\{S_{j_{1}},S_{j_{2}},\ldots,S_{j_{N-n}}\}$, with
$1\leq j_{1}<j_{2}<\ldots <j_{N-n} \leq N$, and $i_{q}\neq j_{p}$.
Let
\begin{equation}
\rho_{A_{n}} \equiv \rho_{i_{1},i_{2},\ldots,i_{n}} \,=\,
Tr_{B_{N-n}}[\rho] \,=\, Tr_{j_{1},j_{2},\ldots,j_{N-n}}[\rho] \,
\label{reducedrho}
\end{equation}
denote the reduced density matrix of subsystem $S_{A_{n}}$ after
tracing over subsystem $S_{B_{N-n}}$. The linear entropy associated
to such a bipartition is defined as
\begin{equation}
S_{L}^{(A_{n};B_{N-n})}(\rho) \,=\, \frac{d}{d-1}(1-Tr_{A_{n}}[\rho_{A_{n}}^{2}]) \,,
\label{linearentropy}
\end{equation}
where the $d$ is the Hilbert-space dimension given by $d=\min\{\dim
S_{A_{n}}\,,\dim S_{B_{N-n}}\}=\min\{2^{n},2^{N-n}\}$. Finally, we
introduce the average linear entropy
\begin{equation}
\langle S_{L}^{(n:N-n)}(\rho) \rangle \,=\, \left(%
\begin{array}{c}
  N \\
  n \\
\end{array}%
\right)^{-1} \; \sum_{A_{n}} S_{L}^{(A_{n};B_{N-n})}(\rho) \,,
\label{avlinearentr}
\end{equation}
where the sum is intended over all the possible bi-partitions of the
system in two subsystems, respectively with $n$ and $N-n$ elements
$(1\leq n <N)$ \cite{Oliveira}.

We can now compute the linear
entropies (\ref{linearentropy}) and (\ref{avlinearentr}) for the
two-qubit Bell state $|\nu_{\alpha}(t)\rangle$, i.e.
Eq.~(\ref{flavorBellstate}), with density matrix
$\rho^{(\alpha)}=|\nu_{\alpha}(t)\rangle\langle\nu_{\alpha}(t)|$.
The linear entropy associated to the reduced state after tracing
over one mode (flavor) can be computed straightforwardly:
\begin{eqnarray}
S_{L \alpha}^{(\mu;e)} \,=\, S_{L \alpha}^{(e;\mu)}
&=& 4 |\mathbf{\widetilde{U}}_{\alpha e}(t)|^{2}
\, |\mathbf{\widetilde{U}}_{\alpha \mu}(t)|^{2}
\nonumber \\
&=& 4
|\mathbf{\widetilde{U}}_{\alpha e}(t)|^{2} \,
(1-|\mathbf{\widetilde{U}}_{\alpha e}(t)|^{2})
\nonumber \\
&=& 4
|\mathbf{\widetilde{U}}_{\alpha \mu}(t)|^{2} \,
(1-|\mathbf{\widetilde{U}}_{\alpha \mu}(t)|^{2}) \,.
\label{SLsinglet}
\end{eqnarray}
In Eq.~(\ref{SLsinglet}) and in the following, we use the notation
$S_{L \alpha}^{(e;\mu)}\equiv S_{L}^{(e;\mu)}(\rho^{(\alpha)})$,
where the subscript $\alpha$ refers to the time-evolved state
(channel), and the superscripts $(e;\mu)$ refer to the considered
modes (flavors). Clearly, for the two-flavor state
(\ref{flavorBellstate}), and in general for any two-qubit system,
symmetry imposes $S_{L \alpha }^{(e;\mu)}=S_{L
\alpha}^{(\mu;e)}=\langle S_{L \alpha }^{(1:1)} \rangle$. Expression
(\ref{SLsinglet}) establishes that the linear entropy of the reduced
state is equal to the product of the two-flavor transition
probabilities. Moreover, for any reduced state $\rho$ of a two-level
system one has that $S_L = 2[1-Tr(\rho^{2})] = 4Det\rho =
4\lambda_1(1-\lambda_1)$, where $\lambda_1$ is one of the two
non-negative eigenvalues of $\rho$, and the relation $\lambda_1 +
\lambda_2 = 1$ has been exploited. Comparing with
Eq.~(\ref{SLsinglet}), one sees that the transition probabilities
coincide
with the eigenvalues of the reduced state density matrix.

\begin{figure}[ht]
\centering
\includegraphics*[width=7.5cm]{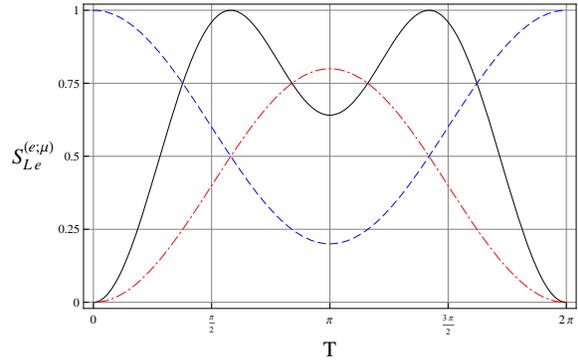}
\caption{(Color online) Linear entropy $S_{L e}^{(e;\mu)}$ (full) as a
function of the scaled time $T=\frac{2Et}{\Delta m_{12}^{2}}$.
The mixing angle $\theta$ is fixed at the experimental
value $\sin^{2}\theta = 0.314$.
The transition probabilities
$P_{\nu_{e}\rightarrow\nu_{e}}$ (dashed) and
$P_{\nu_{e}\rightarrow\nu_{\mu}}$ (dot-dashed) are
reported as well for comparison.}
\label{FigTwoFlav}
\end{figure}

In Fig.~\ref{FigTwoFlav} we show the behavior of $S_{L
e}^{(e;\mu)}$ as a function of the scaled, dimensionless time
$T=\frac{2 E t}{\Delta m_{12}^{2}}$. In the same figure, we also
report the behavior of the transition probabilities
$P_{\nu_{e}\rightarrow\nu_{e}}$ and
$P_{\nu_{e}\rightarrow\nu_{\mu}}$.
The plots have a clear physical interpretation. At time $T=0$, the
entanglement is zero, the global state of the system is factorized,
and the two flavors are not mixed. For $T>0$, flavors start to
oscillate and the entanglement is maximal at largest mixing:
$P_{\nu_{e}\rightarrow\nu_{e}}=P_{\nu_{e}\rightarrow\nu_{\mu}}=0.5$,
and minimum  at $T=\pi$.

\begin{figure}[h]
\centering
\includegraphics*[width=7.5cm]{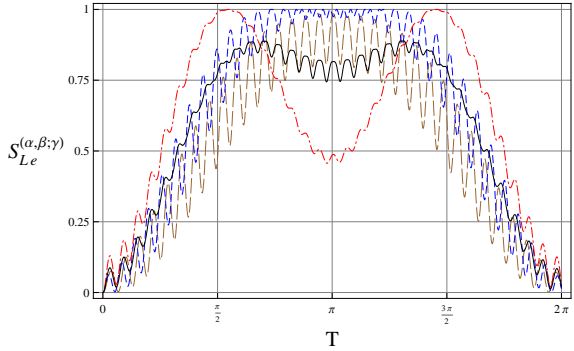}
\caption{(Color online) Linear entropies $S_{L e}^{(\alpha,\beta;\gamma)}$ and
$\langle S_{L e}^{(2;1)}\rangle$ as
functions of the scaled time $T$.
Curves correspond to the partial linear entropies
$S_{L e}^{(e,\mu;\tau)}$ (long-dashed),
$S_{L e}^{(e,\tau;\mu)}$ (dashed),
$S_{L e}^{(\mu,\tau;e)}$ (dot-dashed),
and to the average linear entropy
$\langle S_{L e}^{(2;1)}\rangle$ (full).
Parameters $\theta_{ij}$ and $\Delta m_{ij}^{2}$
are fixed at the central experimental values \cite{Fogli}.}
%%$\sin^{2}\theta_{12}=0.314$, $\sin^{2}\theta_{23}=0.45$,
%%$\sin^{2}\theta_{12}=0.008$,
%%$\Delta m_{12}^{2}=7.92\times 10^{-5} eV^{2}$,
%%$\Delta m_{23}^{2}=2.6\times 10^{-3} eV^{2}$, and
%%uncertainties are neglected.
\label{FigThreeFlav}
\end{figure}

\begin{figure}[ht]
\centering
\includegraphics*[width=7.5cm]{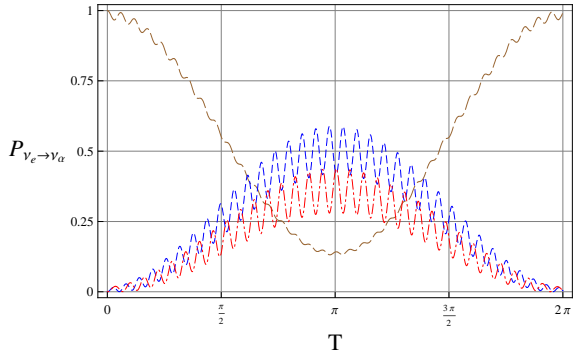}
\caption{(Color online) Transition probabilities $P_{\nu_{e}\rightarrow\nu_{\alpha}}$
as functions of the scaled time $T$. Parameters $\theta_{ij}$ and $\Delta m_{ij}^{2}$
are fixed at the central experimental values as in Fig.~\ref{FigThreeFlav}.
Curves correspond to
$P_{\nu_{e}\rightarrow\nu_{e}}$ (long-dashed),
$P_{\nu_{e}\rightarrow\nu_{\mu}}$ (dashed), and
$P_{\nu_{e}\rightarrow\nu_{\tau}}$ (dot-dashed).}
\label{Fig3}
\end{figure}

 We can now investigate three-flavor oscillations, and the
associated three-qubit $W$-like states (\ref{flavorWstate}).
Similarly to Eq.~(\ref{SLsinglet}), tracing, e. g., over mode
$\tau$, one has
\begin{eqnarray}
S_{L \alpha}^{(e,\mu;\tau)} &=& 4 |\mathbf{\widetilde{U}}_{\alpha
\tau}(t)|^{2} \, (|\mathbf{\widetilde{U}}_{\alpha
e}(t)|^{2}+|\mathbf{\widetilde{U}}_{\alpha
\mu}(t)|^{2}) \nonumber \\
&=& 4 |\mathbf{\widetilde{U}}_{\alpha
\tau}(t)|^{2} \, (1-|\mathbf{\widetilde{U}}_{\alpha \tau}(t)|^{2}) \,.
\label{SLW}
\end{eqnarray}
The linear entropies for the two remaining bi-partitions are easily
obtained by permuting the indexes $e,\mu,\tau$. The average linear
entropy for the state (\ref{flavorWstate}) is then
\begin{eqnarray}
\langle S_{L \alpha}^{(2:1)} \rangle &=&
\frac{8}{3}(|\mathbf{\widetilde{U}}_{\alpha
e}(t)|^{2}|\mathbf{\widetilde{U}}_{\alpha
\mu}(t)|^{2}+|\mathbf{\widetilde{U}}_{\alpha
e}(t)|^{2}|\mathbf{\widetilde{U}}_{\alpha \tau}(t)|^{2} \nonumber \\
&&+|\mathbf{\widetilde{U}}_{\alpha
\mu}(t)|^{2}|\mathbf{\widetilde{U}}_{\alpha \tau}(t)|^{2}) . \label{avSLW}
\end{eqnarray}
%%Relations (\ref{SLW}) and (\ref{avSLW}) can obviously
%%be further generalized to instances involving an arbitrary
%%number of flavors ($N$-qubit, $W$-like states).
In Fig.~\ref{FigThreeFlav}, we show $S_{L
e}^{(\alpha,\beta;\gamma)}$ and $\langle S_{L e}^{(2;1)}\rangle$ as
functions of the scaled time $T=\frac{2Et}{\Delta m_{12}^{2}}$. The
mixing angles $\theta_{ij}$ and the squared mass differences are
fixed at the most recent experimental values reported in
Ref.~\cite{Fogli}.
In order to track the behavior of the entanglement,
we plot in Fig.~\ref{Fig3} the transition probabilities
$P_{\nu_{e}\rightarrow\nu_{\alpha}}$ $(\alpha=e,\mu,\tau)$.
Comparing Fig.~\ref{FigThreeFlav} and  Fig.~\ref{Fig3}, we observe that, as
one may expect, the more mixed are the flavors, the higher is the
global multipartite entanglement of the system. Moreover, the
partial linear entropies $S_{L e}^{(e,\mu;\tau)}$ and $S_{L
e}^{(e,\tau;\mu)}$ measuring the reduced bipartite entanglement,
exhibit a similar behavior due to the strong correlation between the
components $\nu_{\mu}$ and $\nu_{\tau}$. As $T>0$ the probabilities
$P_{\nu_{e}\rightarrow\nu_{\mu}}$ and
$P_{\nu_{e}\rightarrow\nu_{\tau}}$ increase and oscillate while
remaining close. Similar considerations hold for states
$|\nu_{\mu}(t)\rangle$ and $|\nu_{\tau}(t)\rangle$. Entanglement and
flavor transition probabilities for these states exhibit very fast
oscillating behaviors, related to the experimentally measured values
of the mixing parameters.

\begin{figure}[ht]
\centering
\includegraphics*[width=7.5cm]{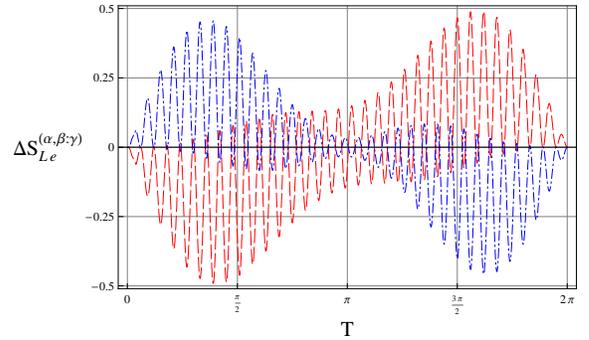}
\caption{(Color online) The imbalances $\Delta S_{L e}^{(\alpha,\beta;\gamma)}$ as
functions of the scaled time $T$.
Curves correspond to
$\Delta S_{L e}^{(e,\mu;\tau)}$ (long-dashed) and
$\Delta S_{L e}^{(e,\tau;\mu)}$ (dot-dashed).
The quantity $\Delta S_{L e}^{(\mu,\tau;e)}$ is vanishing.
Parameters $\theta_{ij}$ and $\Delta m_{ij}^{2}$
are fixed at the central experimental values as in Fig.~\ref{FigThreeFlav}.
The $CP$-violating phase is set at the value $\delta = \pi/2$.}
\label{FigCPviol}
\end{figure}

 Because of $CPT$ invariance, the $CP$ asymmetry
$\Delta_{CP}^{\alpha,\beta}$ is equal to the asymmetry under time
reversal, defined as
\begin{eqnarray}
\Delta_{T}^{\alpha,\beta} &=&
P_{\nu_{\alpha}\rightarrow\nu_{\beta}}(t)-P_{\nu_{\beta}\rightarrow\nu_{\alpha}}(t)
\nonumber \\
&=& P_{\nu_{\alpha}\rightarrow\nu_{\beta}}(t)-P_{\nu_{\alpha}\rightarrow\nu_{\beta}}(-t) \,.
\end{eqnarray}
In the three-flavor instance, such a quantity is different from zero
for a nonvanishing phase $\delta$. It is worth noticing that
$\sum_{\beta}\Delta_{CP}^{\alpha\beta}=0$ with
$\alpha,\beta=e,\mu,\tau$. Introducing the {\it ``imbalances''},
i.e. the difference between the linear entropies and their
time-reversed expressions:
\begin{equation}
\Delta S_{L \lambda}^{(\alpha,\beta;\gamma)} =
S_{L \lambda}^{(\alpha,\beta;\gamma)}(t) -
S_{L \lambda}^{(\alpha,\beta;\gamma)}(-t) \, ,
\label{DeltaSL}
\end{equation}
we can compute, e. g., $\Delta S_{L e}^{(e,\mu;\tau)}$, and obtain:
\begin{equation}
\Delta S_{L e}^{(e,\mu;\tau)} =
4 \Delta_{CP}^{e,\mu} (|\mathbf{\widetilde{U}}_{e\tau}(t)|^{2}
+|\mathbf{\widetilde{U}}_{\tau e}(t)|^{2}-1) \,,
\label{DeltaSL2}
\end{equation}
where the last factor is $CP$-even. In Fig.~\ref{FigCPviol} we show
the behavior of the imbalances $\Delta S_{L
e}^{(\alpha,\beta;\gamma)}$ as functions of time, and see how they
effectively measure $CP$ violation.
%%%%%%%%%%%%%%%%%%%%%%%%%%%%%%%%%%%%%%%%%%%%%%%%%%%%%%%%%%%%%%%%%%%%%%%%%%%%%%%%%%%%%%%%%%%%%%%%%

\begin{figure}[h]
\centering
\includegraphics*[width=6.1cm]{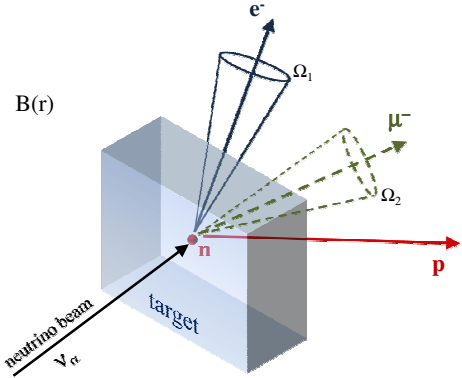}
\caption{(Color online) Scheme for the generation
of single-particle entangled lepton states.
A neutrino beam is focused on a target where events of the following
charged-current interaction may occur:
$\nu_{\alpha} + n \longrightarrow \alpha^{-} + p$ with $\alpha=e\,,\mu$.
The beam is assumed to have energy higher than
the threshold value necessary for the creation of a muon. A spatially
nonuniform magnetic field $\mathbf{B}(\mathbf{r})$ is then applied to
limit the momentum of the outgoing lepton within a certain solid angle $\Omega_{i}$,
and to ensure the spatial separation between the electron and muon spatial paths.
The reaction produces a superposition of electronic and
muonic spatially separated states.}
\label{FigChargCurr}
\end{figure}

In order to demonstrate that the form of single-particle
entanglement encoded in the time-evolved flavor states
$|\underline{\nu}^{(f)}(t)\rangle$ is a real physical resource that
can be legitimately used, at least in principle, for protocols of
quantum information, we discuss an experimental scheme for the
transfer of the flavor entanglement of a neutrino beam into that of
a single-particle system with {\em spatially separated modes}. For
simplicity, we will restrict the analysis to two flavors
$\alpha=e,\mu$. Consider the elementary charged-current interaction
between a neutrino $\nu_{\alpha}$ with flavor $\alpha$ and a nucleon
$N$ \cite{Cheng-Li}. The quasi-elastic scattering interaction yields
the production of a lepton $\alpha^{-}$ and of an outgoing baryon
$X$, according to the reaction:
\begin{equation}
\nu_{\alpha} + N \longrightarrow \alpha^{-} + X \,.
\label{CCinteraction}
\end{equation}
In the simplest instance, the nucleon $N$ is a neutron and the
baryon $X$ is a proton $p$; the corresponding scheme is illustrated
in Fig.~\ref{FigChargCurr}.
Given the initial Bell-like superposition (\ref{flavorBellstate}),
the unitary process associated with the weak interaction
(\ref{CCinteraction}) produces the superposition
\begin{equation}
|\alpha (t)\rangle \,=\, \Lambda_{e}
|1\rangle_{e} |0\rangle_{\mu} + \Lambda_{\mu}
|0\rangle_{e} |1\rangle_{\mu} \,,
\label{fermionBellstate}
\end{equation}
where $|\Lambda_{e}|^{2}+|\Lambda_{\mu}|^{2}=1$, and
$|k\rangle_{\alpha}$, with $k=0,1$, represents the lepton qubit. The
coefficients $\Lambda_{\alpha}$ are proportional to
$\mathbf{\widetilde{U}}_{\alpha \beta}(t)$ and to the cross sections
associated with the creation of an electron or a muon. Comparing our
single-lepton system with the single-photon system, the quantum
uncertainty on ``{\em which path}'' of the photon at the output of
an unbalanced beam splitter is replaced by the uncertainty on ``{\em
which flavor}'' of the produced lepton. The coefficients
$\Lambda_{\alpha}$ play the role of the transmissivity and of the
reflectivity of the beam splitter. Moreover, by exploiting the mass
difference between the two leptons, the desired spatial separation
between the flavors can be achieved by applying a nonuniform
magnetic field. {It is also important to remark that
the approach proposed in the present work can be applied even in
extended neutrino models including one or more sterile neutrinos.
In such cases, from a mathematical point of view the main difference
is that one deals with more than three modes (flavors), while, from an
operational point of view, the presence of sterile neutrinos (undetectable to date)
would introduce a mechanism of loss of quantum information by making the
(observed) mixing matrix non unitary.} Therefore we can conclude that, at least in
principle, the quantum information encoded in the neutrino flavor
states can be transferred to a delocalized two-flavor lepton state,
and the single-particle mode entanglement acquires an operational
characterization that can be exploited for quantum information tasks
using systems of elementary particle physics.

%%However, from a theoretical point of view, a method to test the Bell inequalities
%%for massive single-particle fermion states is still an open problem \cite{Aharonov,Nori}.

%%In conclusion, we have shown that the observable transition probabilities
%%in flavor oscillations essentially coincides with bipartite and
%%multipartite flavor mode entanglement. Therefore, flavor oscillations and
%%flavor mode entanglement are two aspects of the same dynamical phenomenon, i.e.
%%the oscillations can be consistently interpreted as a dynamical redistribution
%%of single-particle entanglement among the different flavor modes. Single-particle
%%entanglement arises as well in particle mixing if the multipartite qubit states
%%are expressed in terms of the mass eigenstates \cite{StaticEntanglement}.\\
%%\indent It is an interesting open question whether such a correspondence
%%persists in a fully quantum field theoretical context, where it
%%has been shown that the mixing of fields with different masses is associated
%%with nontrivial properties of the flavor vacuum \cite{Blasone}.
%%Such a theoretical framework could thus represent a useful playground for
%%the extension of the concept of entanglement to the relativistic domain.

%
\acknowledgments

We acknowledge financial support from MIUR, CNR-INFM Research and Development
Center {\it ``Coherentia''},  INFN,
and from ISI Foundation for Scientific Interchange.

\end{document}